\documentclass[preprint,aps]{revtex4}

\usepackage{graphicx}% Include figure files
\usepackage{dcolumn}% Align table columns on decimal point
\usepackage{bm}% bold math

\font\scripti=cmmi7
\font\scriptscripti=cmmi5
\def\sib#1{\setbox0 = \hbox{\scripti #1}
  \kern-.02em\copy0\kern-\wd0
  \kern.04em\box0} % script italic bold 
\def\ssib#1{\setbox0 = \hbox{\scriptscripti #1}
  \kern-.02em\copy0\kern-\wd0
  \kern.04em\box0} % scriptscript italic bold
\font\tenib=cmmib10 % italic bold for math
\skewchar\tenib='177 \skewchar\tenib='177 \skewchar\tenib='177
\def\bm{\fam10}
\def\pbold#1{\setbox0 = \hbox{$ #1 $}
  \kern-.022em\copy0\kern-\wd0
  \kern.011em\copy0\kern-\wd0
  \kern.011em\copy0\kern-\wd0
  \kern.011em\copy0\kern-\wd0
  \kern.011em\box0} % poorman's bold
  
\def\lesssim{\ \raise.3ex\hbox{$<$}\kern-0.8em\lower.7ex\hbox{$\sim$}\ }
\def\gesim{\ \raise.3ex\hbox{$>$}\kern-0.8em\lower.7ex\hbox{$\sim$}\ }

\begin{document}
\title{A Novel Route to Reach a $p$-wave Superfluid Fermi Gas}
\author{T. Yamaguchi}\author{D. Inotani}\author{Y. Ohashi}
\affiliation{Department of Physics, Keio University, 3-14-1 Hiyoshi, Yokohama 223-8522, Japan}
%%%%%%%%%%%%%%%%%%%%%%%%%%%%%%%%%%%%%%%%%%%%%%%%%%%%%%%%%%%%%%%%%%%%%%%%%%%%%%%
\date{\today}

\begin{abstract}
We theoretically propose an idea to realize a $p$-wave superfluid Fermi gas. To overcome the experimental difficulty that a $p$-wave pairing interaction to form $p$-wave Cooper pairs damages the system before the condensation growth, we first prepare a $p$-wave pair amplitude ($\Phi_p$) in a spin-orbit coupled $s$-wave superfluid Fermi gas, {\it without} any $p$-wave interaction. Then, by suddenly changing the $s$-wave interaction with a $p$-wave one ($U_p$) by using a Feshbach resonance, we reach the $p$-wave superfluid phase with the $p$-wave order parameter being symbolically written as $\Delta_p\sim U_p\Phi_p$. In this letter, we assess this scenario within the framework of a time-dependent Bogoliubov-de Gennes theory. Our results would contribute to the study toward the realization of unconventional pairing states in an ultracold Fermi gas.
\end{abstract}

%\pacs{03.75.Ss, 03.75.-b, 67.85.Lm}
\maketitle
%%%%%%%%%%%%%%%%%%%%%%%%%%%%%%%%%%%%%%%%%%%%%%%%%%%%%%%%%%%%%%%%%%%%%%%%%%%%%%%
\par
In cold Fermi gas physics, one of the most difficult and exciting challenges is the achievement of a $p$-wave superfluid phase transition. Experimentally, one can now tune the strength of a $p$-wave pairing interaction by adjusting the threshold energy of a Feshbach resonance\cite{pwve1,pwve2,pwve3,pwve4,pwve5,pwve6,pwve7}. The formation of $p$-wave molecules by using this interaction has also been reported\cite{pwve4,pwve12,pwve13}. In addition, strong-coupling theories have quantitatively predicted the $p$-wave superfluid phase transition temperature $T_{\rm c}$\cite{Ho,Ohashi,Iskin1,Iskin2,Inotani}. However, the realization of a $p$-wave superfluid Fermi gas has {\it not} been reported yet.
\par
This difficulty comes from the dilemma that, although a $p$-wave pairing interaction is necessary to realize a $p$-wave Fermi gas superfluid, it also damages the system. That is, a $p$-wave interaction causes dipolar relaxations\cite{pwve4}, as well as three-body losses\cite{Gurarie1,Gurarie2,Castin}, leading to very short lifetime of $p$-wave pairs ($=5\sim 20$ ms)\cite{Chevy}. As a result, $p$-wave pairs are destroyed before the $p$-wave condensate enough grows ($=O(100~{\rm ms})$). At the same time, the particle loss from the system destroys the system itself. Since all the current experiments toward the realization of a $p$-wave superfluid Fermi gas are facing these problems, a breakthrough seems necessary.
\par
The purpose of this letter is to propose a possible route to reach a $p$-wave superfluid Fermi gas. To explain our idea, it is convenient to recall that a $p$-wave superfluid Fermi gas is well characterized by the $p$-wave superfluid order parameter, 
\begin{equation}
\Delta_{\sigma,\sigma'}^p({\bm p})=
\sum_{{\sib p}'}U_p({\bm p},{\bm p}')\Phi^p_{\sigma,\sigma'}({\bm p}'),
\label{eq.1}
\end{equation}
where $U_p({\bm p},{\bm p}')$ is a $p$-wave pairing interaction. $\Phi^p_{\sigma,\sigma'}({\bm p}')=\langle c_{{\sib p},\sigma}c_{-{\sib p},\sigma'}\rangle$ is $p$-wave pair amplitude, where $c_{{\sib p},\sigma}$ is the annihilation operator of a Fermi atom with pseudospin $\sigma=\uparrow,\downarrow$, describing two atomic hyperfine states forming Cooper pairs. The current experiments always start from a Fermi gas with a $p$-wave pairing interaction $U_p({\bm p},{\bm p}')$, and then cool down the gas to explore the $p$-wave superfluid phase where the pairing interaction $U_p({\bm p},{\bm p}')$ produces the pair amplitude $\Phi^p_{\sigma,\sigma'}({\bm p}')$, leading to non-vanishing $p$-wave superfluid order parameter $\Delta_{\sigma,\sigma'}^p({\bm p})$ in Eq. (\ref{eq.1}). This very reasonable approach is, however, hampered by the above-mentioned problems associated with the same $p$-wave interaction.
\par
\begin{figure}[t]
\begin{center}
\includegraphics[width=9cm,keepaspectratio]{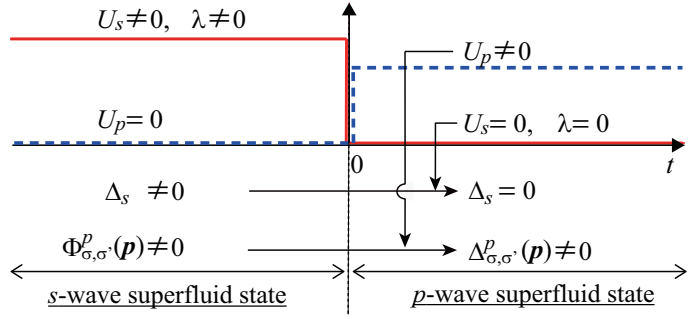}
\caption{(Color online) Schematic picture of our idea to realize a $p$-wave superfluid Fermi gas. When $t<0$, we prepare a $p$-wave pair amplitude $\Phi^p_{\sigma,\sigma'}({\bm p})$ in a spin-orbit coupled $s$-wave superfluid Fermi gas (where $U_s$ and $\lambda$ are an $s$-wave pairing interaction and a spin-orbit coupling, respectively). At this stage, the $p$-wave superfluid order parameter $\Delta_{\sigma,\sigma'}^p({\bm p})$ still vanishes, because of the vanishing $p$-wave interaction, $U_p=0$. At $t=0$, we suddenly replace the $s$-wave interaction by a $p$-wave one ($U_s=0$ and $U_p\ne 0$). The product of the $p$-wave interaction and the prepared $p$-wave pair amplitude $\Phi^p_{\sigma,\sigma'}({\bm p})$ immediately gives non-vanishing $p$-wave order parameter $\Delta_{\sigma,\sigma'}^p({\bm p})$. The $s$-wave order parameter vanishes when $t\ge 0$, because of $U_s=0$. As a result, the $p$-wave superfluid state being characterized by $\Delta_{\sigma,\sigma'}^p$ is realized.
}
\label{fig1}
\end{center}
\end{figure}

The essence of our idea is, in a sense, to inverse this ordinary experimental procedure. That is, as schematically shown in Fig. \ref{fig1}, when $t<0$, we first prepare a Fermi gas with a non-vanishing $p$-wave pair amplitude $\Phi^p_{\sigma,\sigma'}({\bm p})$ {\it without} a $p$-wave pairing interaction ($U_p({\bm p},{\bm p}')=0$). Recently, it has been shown that such preparation is possible in a spin-orbit coupled $s$-wave superfluid Fermi gas\cite{Hu,Yamaguchi}. At this stage, since the $p$-wave interaction $U_p({\bm p},{\bm p}')$ vanishes, one does not suffer from the above-mentioned serious difficulties. Once this first step is accomplished, the $p$-wave superfluid order parameter is then immediately obtained, when one suddenly changes the pairing interaction from the $s$-wave type to an appropriate $p$-wave one ($t\ge 0$ in Fig. \ref{fig1}). 
This manipulation is also possible in cold Fermi gases by adjusting an external magnetic field from an $s$-wave Feshbach resonance to a $p$-wave one. Then, by definition, the system is in the $p$-wave superfluid state with $\Delta^p_{\sigma,\sigma'}({\bm p})$. Just after this manipulation, this order parameter is constructed from the introduced $p$-wave interaction $U_p({\bm p},{\bm p}')$ and the $p$-wave pair amplitude $\Phi^p_{\sigma,\sigma'}({\bm p})$ prepared in advance in the $s$-wave state. Of course, the introduced $p$-wave interaction would start to damage the system as in the ordinary approach. However, the advantage of our approach is that the system is in the $p$-wave superfluid phase, at least just after the $s$-wave interaction is replaced by a $p$-wave one. 
\par
We briefly note that the time evolution of a superfluid Fermi gas after the sudden change of the interaction strength has recently been discussed\cite{t1,t2,t3,t4,t5,t6,t7,t8,t9}. In a sense, our idea is an extension of this so-called quench dynamics, to the case when the pairing symmetry is also suddenly changed. 
\par
%%%%%%%%%%%%%%%%%%%%%%%%%%%%%%%%%%%%%%%%%%%%%%%%%%%%%%%%%%%%%%%%%%%%%%%%%%%%%%%
%%%%%%%%%%%%%%%%%%%%%%%%%%%%%%%%%%%%%%%%%%%%%%%%%%%%%%%%%%%%%%%%%%%%%%%%%%%%%%%
\par
%%%%%%%%%%%%%%%%%%%%%%%%%%%%%%%%%%%%%%%%%%%%%%%%%%%%%%%%%%%%%%%%%%%%%%%%%%%%%%%
\begin{figure}[t]
\begin{center}
\includegraphics[width=5.5cm,keepaspectratio]{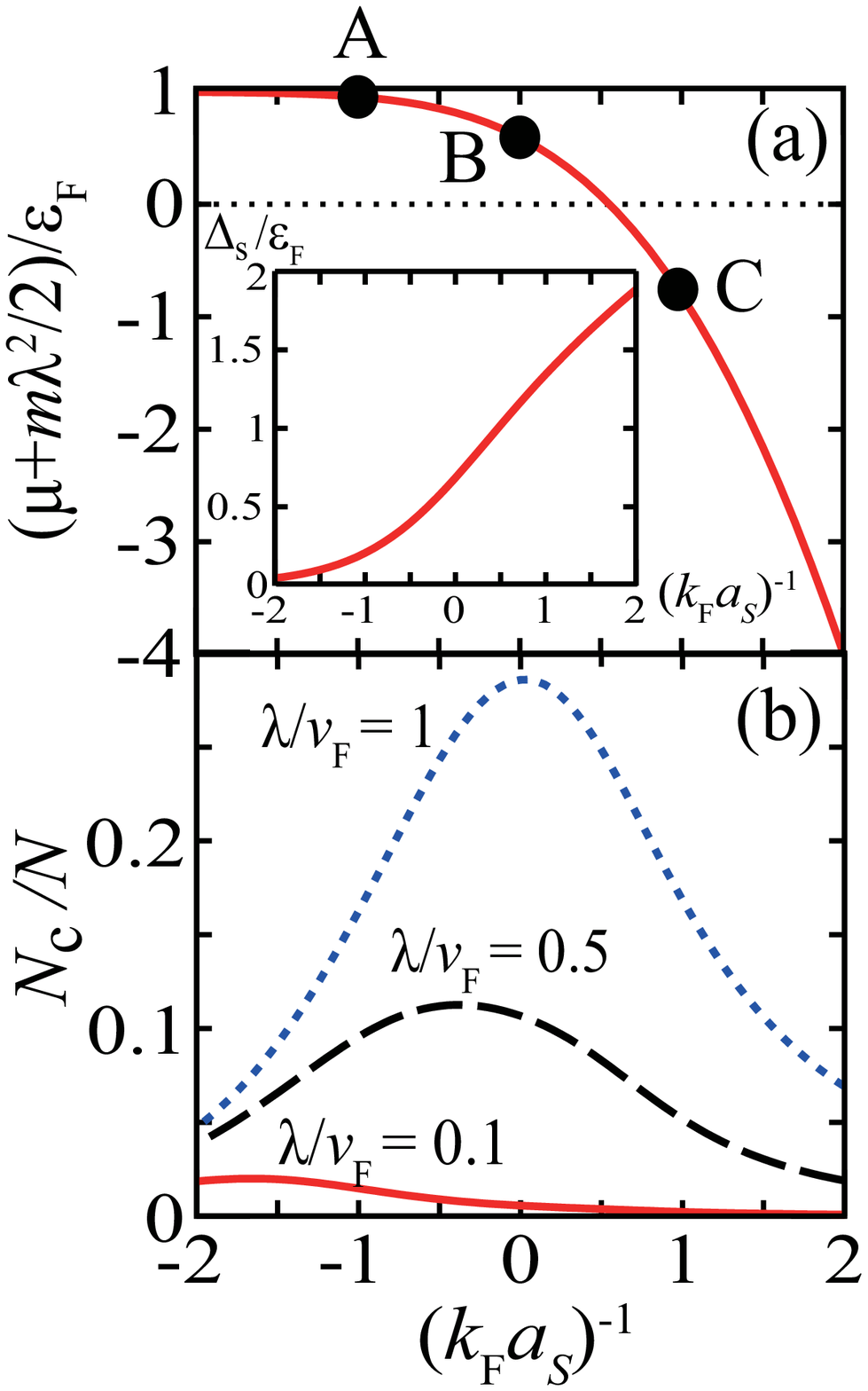}
\caption{(Color online) (a) Calculated Fermi chemical potential $\mu$ in the BCS-BEC crossover regime of a spin-orbit coupled $s$-wave superfluid Fermi gas at $T=0$. When $U_s=0$, $\mu+m\lambda^2/2$ means the Fermi energy measured from the bottom of the energy band. The inset shows the $s$-wave superfluid order parameter $\Delta_s$. We note that $\mu$ and $\Delta_s$ are independent of the spin-orbit interaction $H_{\rm so}=\lambda p_z\tau_x$\cite{Yamaguchi}. (b) $N_{\rm c}\equiv\sum_{\sib p}|\Phi^{\rm t}_{\sigma,\sigma}({\bm p})|^2$. 
A finite value of this quantity means the induction of the triplet pair amplitude $\Phi^{\rm t}_{\sigma,\sigma}({\bm p})$ in the $s$-wave superfluid state. $v_{\rm F}$ is the Fermi velocity and $\varepsilon_{\rm F}\equiv mv_{\rm F}^{2}/2$.
}
\label{fig2}
\end{center}
\end{figure}
%%%%%%%%%%%%%%%%%%%%%%%%%%%%%%%%%%%%%%%%%%%%%%%%%%%%%%%%%%%%%%%%%%%%%%%%%%%%%%%
\par
In assessing our idea, a key is to clarify whether or not the $p$-wave pair amplitude $\Phi_{\sigma,\sigma'}^p({\bm p})$ induced in the $s$-wave state can really produce the $p$-wave superfluid order parameter $\Delta^p_{\sigma,\sigma'}({\bm p})$ after the interaction change. To see this, in this letter, we employ a time-dependent Bogoliubov-de Gennes theory (TDBdG)\cite{timeT1,timeT2}, to examine the time evolution of the $p$-wave order parameter after a $p$-wave interaction is switched on. Although TDBdG cannot deal with the energy relaxation of the system, as well as the particle loss, this approach is still convenient to examine the early behavior of the $p$-wave superfluid after the $p$-wave interaction is turned on. We briefly note that TDBdG has recently been applied to the quench dynamics of a superfluid Fermi gas\cite{t1,t2,t3,t4,t5,t6,t7,t8,t9}. 
\par
To prepare the $p$-wave pair amplitude, we employ a spin-orbit coupled (equilibrium) $s$-wave superfluid Fermi gas\cite{Hu,Yamaguchi}, described by the model Hamiltonian,
\begin{eqnarray}
H_s
&=&
\sum_{{\sib p},\sigma,\sigma'}
\left[
\xi_{\sib p}\delta_{\sigma,\sigma'}
+\lambda p_z\tau_x^{\sigma,\sigma'}
\right]
c_{{\sib p},\sigma}^\dagger 
c_{{\sib p},\sigma'}
\nonumber
\\
&-&U_s\sum_{{\sib p},{\sib p}',{\sib q}}
c_{{\sib p}+{\sib q}/2,\uparrow}^\dagger
c_{-{\sib p}+{\sib q}/2,\downarrow}^\dagger
c_{-{\sib p}'+{\sib q}/2,\downarrow}
c_{{\sib p}'+{\sib q}/2,\uparrow}.
\label{eq.2}
\end{eqnarray}
In this letter, we take $\hbar=1$ and the the system volume $V$ is taken to be unity, for simplicity. In Eq. (\ref{eq.2}), $\xi_{\sib p}=\varepsilon_{\sib p}-\mu={\bm p}^2/(2m)-\mu$ is the kinetic energy of a Fermi atom described by the creation operator $c_{{\sib p},\sigma}^\dagger$, measured from the Fermi chemical potential $\mu$. $-U_s~(<0)$ is a contact-type $s$-wave pairing interaction, which is assumed to be tunable by a Feshbach-resonance\cite{Chin}. As usual in cold Fermi gas physics, we measure the $s$-wave interaction strength in terms of the inverse $s$-wave scattering length $(k_{\rm F}a_s)^{-1}$, scaled by the Fermi momentum $k_{\rm F}$. In this scale, $(k_{\rm F}a_s)^{-1}\lesssim -1$ and $(k_{\rm F}a_s)^{-1}\gesim +1$ correspond to the weak-coupling BCS (Bardeen-Cooper-Schrieffer) regime and the strong-coupling BEC (Bose-Einstein condensation) regime, respectively. The region $-1\lesssim (k_{\rm F}a_s)^{-1}\lesssim +1$ is called the BCS-BEC crossover region. Treating the BCS-BEC  crossover physics at $T=0$ within the framework of the BCS-Leggett theory\cite{Leggett}, we self-consistently determine the chemical potential $\mu$, as well as the $s$-wave superfluid order parameter $\Delta_s=U_s\sum_{\sib p}\langle c_{{\sib p},\uparrow}c_{-{\sib p},\downarrow}\rangle$, from the coupled BCS gap equation,
\begin{equation}
1=-{4\pi a_s \over m}\sum_{\sib p}
\left[
{1 \over 2}\sum_{\alpha=\pm}{1 \over 2E_{\sib p}^\alpha}
-
{1 \over 2\varepsilon_{\sib p}}
\right],
\label{eq.3}
\end{equation}
with the equation for the number $N$ of Fermi atoms,
\begin{equation}
N={1 \over 2}\sum_{{\sib p},\alpha=\pm}
\left[
1-{\xi_{\sib p}^\alpha \over E_{\sib p}^\alpha}
\right].
\label{eq.4}
\end{equation}
Here, $E_{\sib p}^\pm=\sqrt{(\xi_{\sib p}^\alpha)^2+\Delta_s^2}$ (where $\xi_{\sib p}^\alpha=\xi_{\sib p}+\alpha\lambda|p_z|$) describes Bogoliubov single-particle excitations in the presence of the one-component spin-orbit interaction, $H_{\rm so}=\lambda p_z\tau_x^{\sigma,\sigma'}$ (where $\tau_x$ is the Pauli matrix). We briefly note that $H_{\rm so}$ has recently been realized in $^6$Li\cite{Cheuk} and $^{40}$K\cite{Wang} Fermi gases, by using a synthetic gauge field technique.
\par
As pointed out in Refs.\cite{Hu,Yamaguchi}, the spin-orbit interaction $H_{\rm so}$ induces the spin-triplet pair amplitudes in the $s$-wave superfluid state, given by
\begin{equation}
\Phi^{\rm t}_{\uparrow,\uparrow}({\bm p})
=-\Phi^{\rm t}_{\downarrow,\downarrow}({\bm p})=
-{p_z \over |p_z|}\sum_{\alpha=\pm}
{\alpha\Delta_s \over 2E_{\sib p}^\alpha}.
\label{eq.5}
\end{equation}
In Fig. \ref{fig2}, we show the quantity $N_{\rm c}\equiv\sum_{\sib p}|\Phi^{\rm t}_{\sigma,\sigma}({\bm p})|^2$ in the BCS-BEC crossover region, together with $\Delta_s$ and $\mu$ (that are self-consistently determined from the coupled equations (\ref{eq.3}) and (\ref{eq.4})). The induced spin-triplet pair amplitude $\Phi^{\rm t}_{\sigma,\sigma}({\bm p})$ has the $p_z$-wave component $\Phi^{p_z}_{\sigma,\sigma}({\bm p})\propto p_z$\cite{Yamaguchi}, which may also be expected from the prefactor of Eq. (\ref{eq.5}).
\par
To connect the above BCS-Leggett scheme to the TDBdG formalism, it is useful to note that the former result is also obtained as a steady-state solution of the TDBdG equation,
\begin{equation}
{\rm i}\frac{\partial}{\partial t}{\tilde {\bm u}}(t) = {\hat H}_s {\tilde {\bm u}}(t),
\label{eq.6}
\end{equation}
where 
\begin{eqnarray}
{\hat H}_s=
\left(
\begin{array}{cccc}
\varepsilon_{\sib p} & \lambda p_{z} & 0 & {\tilde \Delta}_s(t)  \\
\lambda p_{z} & \varepsilon_{\sib p} & -{\tilde \Delta}_s(t) & 0  \\
0 & -{\tilde \Delta}_s^*(t) & -\varepsilon_{\sib p} & \lambda p_z \\
{\tilde \Delta}_s^*(t) & 0 & \lambda p_z & -\varepsilon_{\sib p} 
\end{array}
\right).
\label{eq.7}
\end{eqnarray}
In Eq. (\ref{eq.6}), ${\tilde {\bm u}}(t)=({\tilde u}_{{\sib p},\uparrow}^\alpha(t), {\tilde u}_{{\sib p},\downarrow}^\alpha(t), {\tilde v}_{{\sib p},\uparrow}^\alpha(t), {\tilde v}_{{\sib p},\downarrow}^\alpha(t))^T$ consists of the coefficients of the Bogoliubov transformation,
\begin{equation}
c_{{\sib p},\sigma}(t) = \sum_\alpha
\left[
{\tilde u}_{{\sib p},\sigma}^\alpha(t)\gamma_{{\sib p},\alpha}
+{\tilde v}_{-{\sib p},\sigma}^{\alpha *}(t)\gamma_{-{\sib p},\alpha}^{\dagger}
\right].
\label{eq.8}
\end{equation}
Here, the normalization condition $|{\tilde u}_{{\sib p},\sigma}^\alpha(t)|^{2}+|{\tilde v}_{{\sib p},\sigma}^\alpha(t)|^2 = 1$ is imposed. The Fermi operator $\gamma_{{\sib p},\alpha}$ describes Bogoliubov excitations. Substituting ${\tilde u}_{{\sib p},\sigma}^\alpha(t)=u_{{\sib p},\sigma}^\alpha e^{-{\rm i}(E+\mu)t}$, ${\tilde v}_{{\sib p},\sigma}^\alpha(t)=v_{{\sib p},\sigma}^\alpha e^{-{\rm i}(E-\mu)t}$, and ${\tilde \Delta}_s(t)=\Delta_s e^{-2{\rm i}\mu t}$ (where $\Delta_s$ is assumed to be real) into Eq. (\ref{eq.6}), we obtain the ordinary (time-independent) BdG equation as,
\begin{eqnarray}
\left(
\begin{array}{cccc}
\xi_{\sib p} & \lambda p_{z} & 0 & \Delta_s  \\
\lambda p_{z} & \xi_{\sib p} & -\Delta_s & 0  \\
0 & -\Delta_s & -\xi_{\sib p} & \lambda p_z \\
\Delta_s & 0 & \lambda p_z & -\xi_{\sib p} 
\end{array}
\right)
{\bm u}_0
=
E
{\bm u}_0,
\label{eq.8b}
\end{eqnarray}
where ${\bm u}_0=(u_{{\sib p},\uparrow}^\alpha, u_{{\sib p},\downarrow}^\alpha, v_{{\sib p},\uparrow}^\alpha, v_{{\sib p},\downarrow}^\alpha)^T$. The $s$-wave order parameter $\Delta_s=-(U_s/2)\sum_{{\sib p},\alpha}u_{{\sib p},\uparrow}^{\alpha}v_{{\sib p},\downarrow}^{\alpha *}$ and the number equation $N=(1/2)\sum_{{\sib p},\sigma,\alpha}|v_{{\sib p},\sigma}^\alpha|^2$ in the BdG formalism reproduce Eqs. (\ref{eq.3}) and (\ref{eq.4}), respectively. The eigen-energy equals $E_{\sib p}^\pm$ given below Eq. (\ref{eq.4}).
\par
We now examine the time-evolution of the system, after the sudden change of the pairing interaction from the $s$-wave type to a $p$-wave one ($t\ge 0$). As a possible situation in an ultracold Fermi gas, we consider the case when an external magnetic field is rapidly tuned from an $s$-wave Feshbach resonance field (which gives $U_s$ in Eq. (\ref{eq.2})) to a $p$-wave Feshbach resonance field of the $\uparrow$-spin component, giving the $p$-wave pairing interaction,\begin{equation}
V_p={1 \over 2}\sum_{{\sib p},{\sib p}',{\sib q}}U_p({\bm p},{\bm p}')c_{{\sib p}+{\sib q}/2,\uparrow}^\dagger c_{-{\sib p}+{\sib q}/2,\uparrow}^\dagger c_{-{\sib p}'+{\sib q}/2,\uparrow} c_{-{\sib p}'+{\sib q}/2,\uparrow},
\label{eq.9}
\end{equation} 
where $U_p({\bm p},{\bm p}')=-4\pi g_p\sum_{L_z=0,\pm 1} F_{\sib p}^*F_{{\sib p}'}Y_{1,L_z}^*({\hat {\bm p}})Y_{1,L_z}({\hat {\bm p}}')$\cite{Ho,Iskin2}, and $Y_{1,L_z}({\hat {\bm p}})$ are spherical harmonics, with ${\hat {\bm p}}={\bm p}/|{\bm p}|$. $F_{\sib p}=|{\bm p}|p_0/[|{\bm p}|^2+p_0^2]$, where $p_0$ is a cutoff momentum, which we take $p_0=10k_{\rm F}$ in this letter\cite{Liao,note}. The coupling constant $g_p$ is related to the $p$-wave scattering volume $v_p$ as,
\begin{equation}
{4\pi v_pp_0^2 \over m}=-
{g_p \over 1-g_p\sum_{\sib p}F_{\sib p}^2/(2\varepsilon_{\sib p})}.
\label{eq.10}
\end{equation}
\par
Among the three $p$-wave Cooper channels ($L_z=0,\pm 1$), the $(L_z=0)$-component only produces a non-vanishing $p$-wave superfluid order parameter $\Delta_{\uparrow,\uparrow}^p({\bm p})$ in the presence of triplet pair amplitude $\Phi_{\uparrow,\uparrow}^{\rm t}({\bm p})$ in Eq. (\ref{eq.5}). (Note that $Y_{1,0}\propto p_z$.) Thus, we only retain this channel\cite{Iskin1}. In addition, we turn off the spin-orbit coupling ($\lambda=0$) when $t\ge 0$, and also assume the absence of any other residual interactions, for simplicity. The TDBdG equation for $t\ge 0$ is then given by Eq. (\ref{eq.6}), where ${\hat H}_s$ is replaced by
\begin{eqnarray}
{\hat H}_p=
\left(
\begin{array}{cccc}
\varepsilon_{\sib p} & 0 & \Delta_{\uparrow,\uparrow}^p({\bm p},t) & 0  \\
0 & \varepsilon_{\sib p} & 0 & 0  \\
\Delta_{\uparrow,\uparrow}^{p*}({\bm p},t) & 0 & -\varepsilon_{\sib p} & 0 \\
0 & 0 & 0 & -\varepsilon_{\sib p} 
\end{array}
\right).
\label{eq.11}
\end{eqnarray}
The initial condition is given by ${\tilde {\bm u}}(0)={\bm u}_0$, where ${\bm u}_0$ is obtained from the BdG equation (\ref{eq.8b}) for the $s$-wave superfluid state. The $p$-wave superfluid order parameter $\Delta_{\uparrow,\uparrow}^p({\bm p},t)=F_{\bm p}^*Y_{1,0}^*({\hat {\bm p}})\Delta_p(t)$ in Eq. (\ref{eq.11}) is calculated from
\begin{equation}
\Delta_p(t)=-2\pi g_p\sum_{\sib p}F_{\sib p}Y_{1,0}({\hat {\bm p}})
\sum_{\alpha=\pm} {\tilde u}_{{\sib p},\uparrow}^\alpha(t){\tilde v}_{{\sib p},\uparrow}^{\alpha*}(t).
\label{eq.12}
\end{equation}
At $t=0$, the $p$-wave superfluid order parameter $\Delta_p(t=0)$ is evaluated by using ${\bm u}_0$ prepared in the $s$-wave superfluid state. Since the $\uparrow$-spin component is completely disconnected from the $\downarrow$-spin component when $t\ge 0$, one may only deal with the $\uparrow$-spin component in the TDBdG equation (\ref{eq.6}), when $t\ge 0$.
\par
%%%%%%%%%%%%%%%%%%%%%%%%%%%%%%%%%%%%%%%%%%%%%%%%%%%%%%%%%%%%%%%%%%%%%%%%%%%%%%%
\begin{figure}[t]
\begin{center}
\includegraphics[width=5.5cm,keepaspectratio]{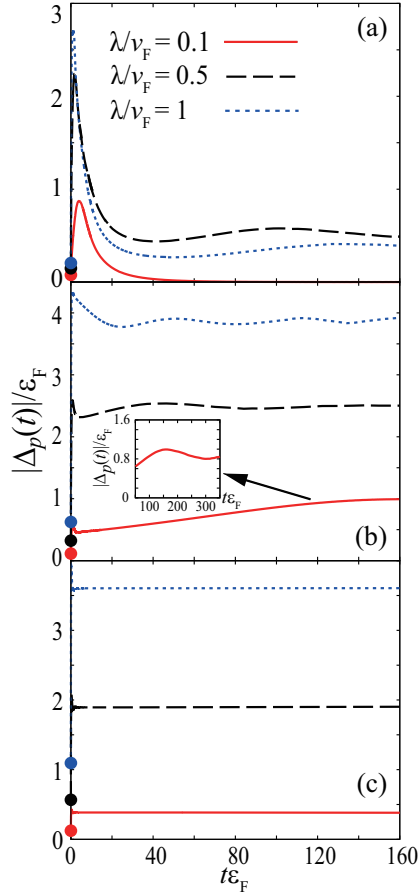}
\caption{(Color online) Time evolution of the $p$-wave superfluid order parameter $\Delta_p(t\ge 0)$, after the $s$-wave pairing interaction is suddenly replaced by the $p$-wave one $V_p$ in Eq. (\ref{eq.9}). (a) A$\to$D. (b) B$\to$E. (c) C$\to$F. Here, (A, B, C) and (D, E, F) represent the $s$-wave interaction strengths and the $p$-wave interaction strengths shown in Figs. \ref{fig2} and \ref{fig4}, respectively. In each case, the filled circle at $t=0$ shows the initial value $|\Delta_p(t=0)|$. The values of $\lambda$ are the spin-orbit coupling strengths when $t<0$. (Note that $\lambda=0$ when $t\ge 0$.) In panel (b), although $|\Delta_p(t)|$ looks increasing with the time evolution when $\lambda/v_{\rm F}=0.1$, it actually oscillates as shown in the inset.
}
\label{fig3}
\end{center}
\end{figure}
%%%%%%%%%%%%%%%%%%%%%%%%%%%%%%%%%%%%%%%%%%%%%%%%%%%%%%%%%%%%%%%%%%%%%%%%%%%%%%%
\par
%%%%%%%%%%%%%%%%%%%%%%%%%%%%%%%%%%%%%%%%%%%%%%%%%%%%%%%%%%%%%%%%%%%%%%%%%%%%%%%
\begin{figure}[t]
\begin{center}
\includegraphics[width=5.5cm,keepaspectratio]{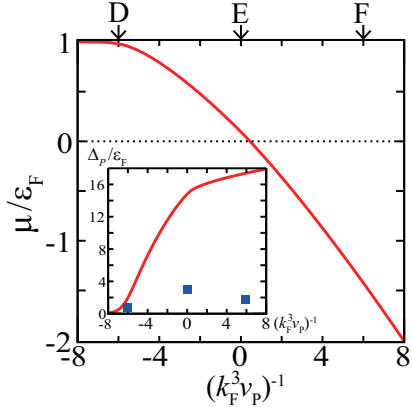}
\caption{(Color online) Calculated Fermi chemical potential $\mu$ in the $p$-wave BCS ground state with the pairing interaction $V_p$ in Eq. (\ref{eq.9}). The inset shows the $p$-wave superfluid order parameter $\Delta_p$. These are obtained within the framework of the BCS-Leggett theory. In the inset, the solid squares are $|\Delta_p(t\varepsilon_{\rm F}=100)|$ in Fig. \ref{fig3} when $\lambda/v_{\rm F}=0.5$.
}
\label{fig4}
\end{center}
\end{figure}
%%%%%%%%%%%%%%%%%%%%%%%%%%%%%%%%%%%%%%%%%%%%%%%%%%%%%%%%%%%%%%%%%%%%%%%%%%%%%%%
\par
Figure \ref{fig3} shows the time evolution of the $p$-wave order parameter $\Delta_p(t\ge 0)$. In this figure, the expected non-vanishing $p$-wave superfluid order parameter is found to really be obtained at $t=0$ from the triplet pair amplitude $\Phi^{\rm t}_{\uparrow,\uparrow}({\bm p})$ which has already been induced in the $s$-wave superfluid state (solid circles in Fig. \ref{fig3}). When the interaction is tuned from the $s$-wave weak-coupling BCS regime (``A" in Fig. \ref{fig2}, where $\mu+m\lambda^2/2\sim\varepsilon_{\rm F}$) to the $p$-wave weak-coupling regime (``D" in Fig. \ref{fig4}, where $\mu\sim\varepsilon_{\rm F}$), Fig. \ref{fig3}(a) shows that, although $\Delta_p(t)$ initially increases with the time evolution, it decreases to vanish when $\lambda/v_{\rm F}=0.1$. Thus, our idea does not work in this case. On the other hand, $\Delta_p(t)$ survives even at $t\varepsilon_{\rm F}=160$ for $\lambda/v_{\rm F}=0.5,~1$. When we use the typical value $\varepsilon_{\rm F}=1$~$\mu$K in an ultracold Fermi gas, one finds that $t\varepsilon_{\rm F}\sim 10^{-2}$ ms. Thus, these two cases can be used to produce a $p$-wave superfluid Fermi gas within the short lifetime of $p$-wave pairs ($=5\sim 20$ ms\cite{Chevy}).
\par
Figures \ref{fig3}(b) and (c), respectively, show the cases when the interaction is tuned from the $s$-wave intermediate-coupling regime (``B" in Fig. \ref{fig2}, where $(k_{\rm F}a_s)^{-1}=0$) to the $p$-wave intermediate-coupling regime (``E" in Fig. \ref{fig4}, where $\mu\sim 0$), and the $s$-wave strong-coupling BEC regime (``C" in Fig. \ref{fig2}, where $\mu+m\lambda^2/2<0$) to the $p$-wave strong-coupling regime (``F" in Fig. \ref{fig4}, where $\mu<0$). These results confirm the validity of our idea, that is, apart from the oscillating behaviors seen in panel (b), $\Delta_p(t)$ approaches a nearly constant value within shorter time scale than the lifetime of $p$-wave pairs.
\par
%%%%%%%%%%%%%%%%%%%%%%%%%%%%%%%%%%%%%%%%%%%%%%%%%%%%%%%%%%%%%%%%%%%%%%%%%%%%%%%
\begin{figure}[t]
\begin{center}
\includegraphics[width=5.5cm,keepaspectratio]{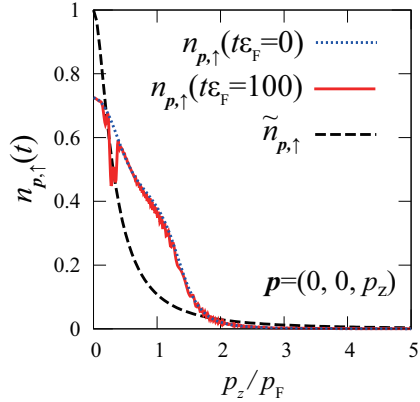}
\caption{(Color online) Calculated momentum distribution $n_{{\sib p},\uparrow}(t)=(1/2)\sum_{\alpha}|\tilde{v}_{{\sib p},\uparrow}^\alpha(t)|^2$ in the case of Fig. \ref{fig3}(b) when $\lambda/v_{\rm F}=0.5$. ${\tilde n}_{{\sib p},\uparrow}$ is the momentum distribution in the $p$-wave BCS ground state. 
}
\label{fig5}
\end{center}
\end{figure}
%%%%%%%%%%%%%%%%%%%%%%%%%%%%%%%%%%%%%%%%%%%%%%%%%%%%%%%%%%%%%%%%%%%%%%%%%%%%%%%
\par
We note that, although non-vanishing $\Delta_p(t)$ is obtained, it is smaller than that expected in the $p$-wave BCS ground state (see the inset in Fig. \ref{fig4}). This is because TDBdG ignores the relaxation of the system into the ground state. Indeed, as shown in Fig. \ref{fig5}, the momentum distribution $n_{{\sib p},\uparrow}(t)=(1/2)\sum_{\alpha}|\tilde{v}_{{\sib p},\uparrow}^\alpha(t)|^2$ is different from that in the $p$-wave BCS ground state, even at $t\varepsilon_{\rm F}=100$. Instead, $n_{{\sib p},\uparrow}(t\varepsilon_{\rm F}=100)$ is rather close to the initial distribution $n_{{\sib p},\uparrow}(t=0)$, which is just the momentum distribution in the $s$-wave superfluid state when $t<0$.
\par
To summarize, we have discussed a possible idea to realize a $p$-wave superfluid Fermi gas. To overcome the difficulty that a $p$-wave pairing interaction damages the system before the enough growth of $p$-wave condensate, our idea first prepares a $p$-wave pair amplitude in a spin-orbit coupled $s$-wave superfluid Fermi gas, without relying on any $p$-wave interactions. Then, by suddenly change the pairing interaction from the $s$-wave type to an appropriate $p$-wave one, we can immediately reach the $p$-wave superfluid phase, at least just after this manipulation. In this letter, we numerically assessed this scenario within the analyses by a time-dependent Bogoliubov-de Gennes equation (TDBdG) at $T=0$.
\par
Since TDBdG cannot treat the energy relaxation of the system into the ground state, the $p$-wave superfluid phase obtained in this letter is different from the  equilibrium BCS ground state. Although it is an interesting future problem to include this relaxation effect, since the magnitude of the $p$-wave superfluid order parameter obtained in this letter is smaller than the equilibrium mean-field BCS value (see the inset in Fig. \ref{fig4}), this effect is expected to enhance $\Delta_p(t)$ compared to the present result. 
\par
Including the lifetime of $p$-wave pairs is also another crucial future challenge, in order to assess our proposal quantitatively. However, at least qualitatively, TDBdG analyses confirmed the possibility that one may produce $p$-wave superfluid order parameter without suffering from the serious lifetime effect that all the current experiments are facing, our results would be still important for the research toward the realization of unconventional pairing states in an ultracold Fermi gas.
\par
In this letter, we have focused on the assessment of our idea, so that we did not discuss detailed physical properties of the $p$-wave state at $t\ge 0$, such as the oscillation of $\Delta_p(t)$ seen in Fig. \ref{fig3}, as well as the dip structure of $n_{{\sib p},\uparrow}(t\varepsilon_{\rm F}=100)$ seen in Fig. \ref{fig5}. Since our idea does not always work as shown in Fig. \ref{fig3}(a), we should also clarify the condition for this idea. We will discuss these issues in our future papers.   
\par
%%%%%%%%%%%%%%%%%%%%%%%%%%%%%%%%%%%%%%%%%%%%%%%%%%%%%%%%%%%%%%%%%%%%%%%%%%%%%%%
\begin{acknowledgments}
This work was supported by KiPAS project in Keio University, as well as Grant-in-aid for Scientific Research from MEXT and JSPS in Japan (No.JP16K17773, No.JP15H00840, No.JP15K00178, No.JP16K05503). 
\end{acknowledgments}
%%%%%%%%%%%%%%%%%%%%%%%%%%%%%%%%%%%%%%%%%%%%%%%%%%%%%%%%%%%%%%%%%%%%%%%%%%%%%%%
\newpage
%%%%%%%%%%%%%%%%%%%%%%%%%%%%%%%%%%%%%%%%%%%%%%%%%%%%%%%%%%%%%%%%%%%%%%%%%%%%%%%
\par


\begin{thebibliography}{99}
\bibitem{note} Choosing this value of $p_0$ is simply due to computational problems. The cutoff-momentum $p_0$ is related to the effective range $p_{\rm eff}$ as\cite{Ho,Iskin2}, $p_{\rm eff}= -(\pi p_0^2/m^2)\sum_{\sib p}F_{\sib p}^2/\varepsilon_{\sib p}^2-2/(p_0^2v_p)$. In a $^{40}$K Fermi gas\cite{pwve2}, as well as in a $^6$Li Fermi gas\cite{pwve7}, $p_{\rm eff}$ has been estimated as $p_{\rm eff}\cong -0.05 a_{\rm B}^{-1} \sim -0.02 a_{\rm B}^{-1}$ near a $p$-wave Feshbach resonance (where $a_{\rm B}$ is the Bohr radius). This gives $p_0\sim 10^2k_{\rm F}\gg 10k_{\rm F}$ for the typical value of the Fermi momentum, $k_{\rm F}^{-1}\sim 100$ nm. Although our smaller $p_0=10k_{\rm F}$ is expected to qualitatively not affect our results, $|\Delta_p(t)|$ is considered to be underestimated\cite{Iskin1}.
\end{thebibliography}
\end{document}